\documentclass[floats,aps,amsmath,amssymb,prl,twocolumn,tightenlines]{revtex4}

\def\dd{{\rm d}}\def\ee{{\rm e}}\def\ii{{\rm i}}
\def\half{{\textstyle{\frac12}}}
\def\beq{\begin{equation}}\def\eeq{\end{equation}}
\def\bea{\begin{eqnarray}}\def\eea{\end{eqnarray}}

\begin{document}

\title{Link between quantum measurement\\ and the $\ii\epsilon$ term in the QFT propagator}

\author{Roman Sverdlov\footnote{roman@phy.olemiss.edu}}
\author{Luca Bombelli\footnote{luca@phy.olemiss.edu}}
\affiliation{Department of Physics and Astronomy, 108 Lewis Hall, University of Mississippi, University, MS 38677, USA}

\date{November 15, 2014}
\begin{abstract}
\noindent Mensky has suggested to account for ``continuous measurement" by attaching to a path integral a weight function centered around the classical path that the integral assigns a probability amplitude \emph{to}. We show that in fact this weight function doesn't have to be viewed as an additional ingredient put in by hand. It can be derived instead from the conventional path integral if the infinitesimal term $\ii\epsilon$ in the propagator is made finite; the ``classical trajectory" is proportional to the current. 
\end{abstract}

\maketitle

\subsection*{1. Introduction}

\noindent In quantum theory, when calculating probabilities of finding a given set of outcomes for a series of observations on a system, one goes through calculations in which the measuring apparatus plays no role and the system is effectively treated as an open one. One possible way to include the measuring apparatus was proposed by Mensky \cite{Proc,Mensky,Kent}, who includes a new ingredient, a weight function $w$, in the definition of the path integral for the probability amplitude of a ``classical" history \footnote{Here, by classical function we simply mean a well-defined field, rather than a solution of the field equation.}. Using a scalar field $\phi$ (for definiteness), the weighted path integral associated with the classical field $\phi_{\rm cl}$ is given by
\beq
Z_{\rm M}(\phi_{\rm cl};w)
= \int [{\cal D}\phi]\, \ee^{\ii S(\phi)}\, w(\phi,\phi_{\rm cl})\;, \label{Mensky1}
\eeq
where $S(\phi)$ is the action. In Mensky's proposal the width of the weight function $w$ is related to the size of the laboratory or the measuring instruments.

We will adopt Mensky's formalism, but with a different interpretation: the weight function will be fundamental rather than just a convenient tool, and its width will accordingly be a constant of nature, whose value determines the separation between the classical and quantum scales for fluctuations around $\phi_{\rm cl}$, rather than the size of the laboratory. In other words, a quantum system is being intrinsically measured continuously in time with a constant ``strength". At quantum scales, the weight function is approximately constant over a very wide range of histories, and leaves the amplitudes associated with quantum fluctuations unchanged. On a classical scale, however, because of the finite width of $w$ we are effectively integrating over a very narrow set of paths (Mensky's ``corridor" \cite{Mensky}), all of which approximate the same classical trajectory. This distinction modifies the standard ontology of quantum theory, along the lines of Kent's proposal of a weighted path integral \cite{Kent}, and implies that, when a wave function splits between macroscopically distinct branches, only one of those branches fits within each corridor, and the probabilities calculated from Eq.\ (\ref{Mensky1}) tell us which branch is selected.

In this paper, we will see how this point of view is related to an interpretation of the $\ii\epsilon$ term in the conventional propagator for a scalar field \cite{PS},
\beq
\int \dd^4p\, \frac{\ee^{\ii p\cdot x}}{p^2-m^2-\ii\epsilon}\;, 
\label{ConvPropagator}
\eeq
and show that taking the weight function to be a Gaussian,
\beq
w(\phi, \phi_{\rm cl}) = \exp\!\bigg(- \frac{\alpha}{2} \int \dd^4 x \;
(\phi (x) - \phi_{\rm cl} (x))^2 \bigg), \label{Gaussian}
\eeq
with a constant, finite rather than infinite corridor width $\alpha^{-1/2}$, is equivalent to taking $\epsilon$ in Eq. \ref{ConvPropagator} to be small but finite rather than infinitesimal, with a specific physical interpretation, and making the sink/source $J$ in the conventional Lagrangian
\beq
{\cal L} = \half\,\partial^\mu\phi\,\partial_\mu\phi + \frac{m^2}{2}\,\phi^2 - J\phi \label{ConvLagrangian}
\eeq
proportional to $\phi_{\rm cl}$. In particular, this equivalence implies that if we take $\epsilon$ seriously rather than simply considering it as a trick to get a well-defined propagator, we can derive the Mensky formalism with a Gaussian form for the weight function, which does not have to be put in by hand. We will show explicitly the steps in one direction, in particular that taking $\epsilon$ to have a finite value leads to a finite corridor width; the steps to show the opposite implication proceed in a very similar way.

\subsection*{2. Calculations}

\noindent Since we are taking the $\ii\epsilon$ term in the propagator seriously, we have to make the corresponding modifications to the Lagrangian. It is easy to see that the presence of the $\ii\epsilon$ term in the propagator can be interpreted as identifying the square of the physical mass with $m^2 - \ii\epsilon$, and that the same identification will result in the Lagrangian
\beq
{\cal L} = \half\, \partial^\mu\phi\, \partial_\mu\phi + \frac{m^2-\ii\epsilon}{2} \,\phi^2 - J \phi\;. \label{Lag}
\eeq

Consider now the two-point function, given by 
\beq
D(x_1, x_2) = \frac{\partial^2 \log Z}{\partial J(x_1)\,\partial J(x_2)}\bigg\vert_{J(x)=0}, \label{2pt}
\eeq
where the partition function $Z(J) = \int[{\cal D}\phi]\,\ee^{\ii S(J,\phi)}$, with $S(J,\phi) = \int\dd^4x\,{\cal L}$, is given by \cite{PS}
\bea
& &Z(J) = \int [{\cal D} \phi] \exp\bigg(\ii \int \dd^4x
\bigg(\half\,\partial^\mu\phi\, \partial_\mu\phi\ +\kern20pt\nonumber\\
& &\kern90pt+\ \frac{m^2-\ii\epsilon}{2}\, \phi^2 - J\phi \bigg) \bigg). \label{Z}
\eea
As long as $x_1 \neq x_2$, the function $D (x_1,x_2)$ will remain the same if we replace $\cal L$ by ${\cal L}'(J,\phi) := {\cal L} (J,\phi) + f(J)$, with $f$ an arbitrary function of $J$:
\bea
& &\frac{\partial^2}{\partial J(x_1)\, \partial J(x_2)} \log Z'\nonumber\\
& &= \frac{\partial^2}{\partial J(x_1)\, \partial J(x_2)}\,
\bigg\{\log \int {\cal D}\phi \times\nonumber\\
& &\kern25pt\times \exp\!\bigg( \int \dd^4x \; \big({\cal L}(\phi, J; x)
+ f(J(x))\big) \bigg) \bigg\}\nonumber\\
& &= \frac{\partial^2}{\partial J(x_1)\, \partial J(x_2)} \log \bigg[ \exp\! \bigg( \int \dd^4x \; f(J(x)) \bigg) \times\nonumber\\
& &\kern55pt\times \int {\cal D} \phi \; \exp\!\bigg(\int
\dd^4x \; {\cal L} (\phi, J, x) \bigg) \bigg] \nonumber\\
& &= \frac{\partial^2}{\partial J(x_1)\,\partial J(x_2)}\bigg(\int\dd^4x\, f(J(x))
+ \log Z\bigg) \nonumber\\
& &= \frac{\partial^2}{\partial J(x_1)\, \partial J(x_2)} \log Z\;. \label{Zprime}
\eea
Now, if in particular we select
\beq
f(J) =  \frac{\ii J^2}{2 \epsilon}
\eeq
and add this to (\ref{Lag}) we obtain a complete square,
\bea
& &{\cal L} + \frac{\ii J^2}{2 \epsilon}
= \half\, \partial^\mu\phi\,\partial_\mu\phi + \frac{m^2 - \ii\epsilon}{2}\, \phi^2 - J\phi + \frac{\ii J^2}{2 \epsilon}\nonumber\\
& &= \half\, \partial^\mu\phi\,\partial_\mu\phi + \frac{m^2}{2}\,\phi^2 - \frac{\ii\epsilon}{2} \left(\phi - \frac{\ii J}{\epsilon}\right)^{\!2}.
\eea
It is easy to see that the partition function $Z'(\phi)$ we get from this Lagrangian density matches the path integral obtained using the $w$ from (\ref{Gaussian}) in (\ref{Mensky1}),
\bea
& &Z_{\rm M}(\phi_{\rm cl};w) = \int[{\cal D} \phi]\bigg[\exp\!\bigg(\ii\int\dd^4x
\bigg(\half\,\partial^\mu\phi\,\partial_\mu\phi + \nonumber\\
& &+\ \frac{m^2}{2} \phi^2 \bigg) \bigg) \exp\!\bigg( - \frac{\alpha}{2} \int\dd^4x\, (\phi - \phi_{\rm cl})^2 \bigg)\bigg], \label{Mensky2}
\eea
if we make the identifications
\beq
\epsilon = \alpha\; , \quad J = - \ii\epsilon\, \phi_{\rm cl}\;. \label{J}
\eeq
Notice that in order for this match to work we have to keep $\phi_{\rm cl}$ real, but allow $J$ to become imaginary. The latter fact will only affect the overall sign ($\ii^2 = -1$) of the propagator, while its magnitude will remain unchanged.

One point should be clarified here. The reader may wonder how we went from considering $J(x) = 0$, for example in Eq.\ \ref{2pt}, to $J = - \ii\epsilon\, \phi_{\rm cl} \ne 0$. The answer is that our goal is to evaluate probability amplitudes for non-zero classical fields as path integrals around the corresponding non-zero functions $\phi_{\rm cl}$, but those can be obtained as Taylor series expansions
\bea
& &\kern-14pt Z(\phi_{\rm cl}) = \sum_n\bigg(\frac{\partial}{\partial J(x_1)}\cdots
\frac{\partial}{\partial J(x_n)} \int[{\cal D}\phi]\,\ee^{\ii S(\phi)}\bigg)_{J = 0}\nonumber \\
& &\kern40pt\times J(x_1)\cdots J(x_n)\;,
\eea
where the integral contains $J$, as can be seen in (\ref{Z}), and all derivatives are evaluated at $J(x) = 0$. Physically, this tells us that a multiparticle system like a large classical object can be obtained as a superposition of terms coming from the interactions among the particles, because the coefficients in the expansion represent the $n$-point functions of the theory.

\subsection*{3. Attenuation of the particle}

\noindent Having established the equivalence between two ways of modifying the calculation of path integrals, we will now see that one consequence of this change is a qualitatively new feature of the theory. Specifically, the fact that the ``mass" has an imaginary contribution coming from the i$\epsilon$ term implies that the exponent of $\ee^{\pm \ii\omega t}$ acquires a \emph{real} term, which makes the magnitude of the exponential deviate from unity. For example, if $t>0$, then when integrating over $\omega$ in expressions like (\ref{ConvPropagator}) we close the contour in the lower half of the complex plane, obtaining
\bea
& &\int \dd \omega \; \frac{\ee^{-\ii\omega t}}{\omega^2 - \omega_k^2 - \ii \epsilon} = \frac{\pi\ii}{\sqrt{\omega_k^2 + \ii \epsilon}}\, \ee^{\ii t \sqrt{\omega_k^2 + \ii \epsilon}}\nonumber\\
& &= \frac{\pi\ii}{\sqrt{\omega_k^2 + \ii \epsilon}}\, \ee^{\ii\omega_k t} \ee^{- \epsilon t/ 2 \omega_k} + O (\epsilon^2)\;. \label{future}
\eea
If, on the other hand, $t<0$, then we close the contour in the upper half, obtaining 
\bea
& &\int \dd \omega \; \frac{\ee^{-\ii\omega t}}{\omega^2 - \omega_k^2 - \ii \epsilon} =  \frac{\pi\ii}{\sqrt{\omega_k^2 + \ii \epsilon}} \ee^{- \ii t \sqrt{\omega_k^2 + \ii \epsilon}}\nonumber\\
& &= \frac{\pi\ii}{\sqrt{\omega_k^2 + \ii \epsilon}} \ee^{-\ii\omega_k t} \ee^{\epsilon t/ 2\omega_k} + O (\epsilon^2)\;. \label{past}
\eea
The $t<0$ and $t>0$ results can be combined into 
\beq
\int \dd\omega \; \frac{\ee^{-\ii\omega t}}{\omega^2 - \omega_k^2 - \ii\epsilon} = \frac{\pi\ii}{\sqrt{\omega_k^2 + \ii\epsilon}}\, \ee^{\ii\omega_k \vert t \vert}\, \ee^{-\epsilon |t|/ 2 \omega_k}\;, \label{BothDirectionsDecay}
\eeq
where the coefficient $\ee^{- \epsilon |t|/ 2 \omega_k}$ indicates that in this framework lifetimes of particles are finite \emph{even if} the particles do not have interactions that would cause them to decay. More specifically, Eq.\ (\ref{past}) means that particles that exist right now were created a finite time in the past, while (\ref{future}) means that those particles will disappear in a finite time. Notice that, because of the time-translation invariance of the theory, particles will appear and disappear at the same rate around any given reference moment in time. This can be understood from Mensky's perspective in the following way. In order for the weight factor to be non-zero, we would like in general the quantity $\phi - \phi_{\rm cl}$ to go to zero at both $t \rightarrow - \infty$ and $t \rightarrow \infty$, while it can become larger in some region of bounded length in between. The extent of this time interval corresponds physically to the lifetime of the particle(s), which the formalism forces to have a finite value $\tau \propto \omega_k/\epsilon$.

This conclusion may sound very surprising, but we should notice that, on a classical scale, due to the very small size of the corridor, the principle of least action approximately holds, and energy-momentum is approximately conserved \emph{at all times}. This is possible because the particles ``disappearing into nothing" are compensated by particles ``created out of nothing" and, although there is no phase correlation between those that disappeared and the newly created ones, we will not notice any difference in experiments involving a large number of particles. The only experiments whose outcomes might be affected are ones involving only a few particles, such as EPR-type experiments, in which we keep track of the relationship between initial and final states in detail, when carried out over very large time scales. In practice, however, we only carry out such experiments on small scales; the detection of cosmological neutrinos, for example, is not a counterexample because we do not observe the initial state and, as far as we know, the neutrinos we see could have been created along the way, replacing other neutrinos emitted by the source. Approximate information about the old particles was stored in the classical field, from which new particles were created, but this information only captures large-scale correlations. The thermodynamic aspects of these processes may be worth studying, but we leave them for future work.

Notice that, in this proposal, the fact that particles appear and disappear is analogous to the effect of sources and sinks in ordinary quantum field theory. The difference with the ordinary theory is that here the sources and sinks are not being viewed as standing for some implicit interactions, nor are they considered as localized $\delta$ functions. In this connection, recall that in Eq.\ \ref{J} we identified $J$, which in the conventional theory destroys the particles, with $-\ii\epsilon\,\phi_{\rm cl}$, where $\phi_{\rm cl}$ is non-material in the sense that it is not a variable integrated over in a path integral, and its effect is spread out over all $t$.

\subsection*{4. Particle lifetimes and the value of $\epsilon$}

\noindent The finite lifetime of a particle can be interpreted as a consequence of continuous measurement (or ``continuous sink"), for which the classical trajectory $\phi_{\rm cl}$ is the measurement outcome, while 
the ``quantum particle" is the fluctuation away from that outcome. If the 
continuous measurement was strong, then the trajectory would be stationary due to the Zeno effect -- which would amount to the particle having a very short lifetime. On the other hand, if the measurement is weak then on time scales much shorter than the ``lifetime" of a particle we don't sense the effects of said measurement, yet on time scales much larger than the lifetime we do sense it, and the particle ``doesn't survive" as we know it. From this point of view, an upper bound on the lifetime of the particle can be determined from the fact that we would like the ``measurement" to be ``strong enough" to account for the collapse of the wave function and the classical behavior of macroscopic objects.

In order for our approach to consistently account for measurement in quantum field theory, we would like this upper bound on $\tau$ to be higher than the lower bound coming from few-particle experiments. If it turned out that this is not the case, however, we could ``redeem" our theory by supplying some extra theory of measurement. One such ``theory of measurement" is the mathematical conjecture that unitary evolution of a wave function would cause it to split into branches that would no longer overlap. If that were the case, then two separate measurement outcomes could co-exist for an arbitrarily long time, until finally one of them disappears due to $\epsilon$. That would ultimately set the upper bound on the lifetime of the particle (which is the same as the time needed for one of the branches to disappear) to infinity, thus removing the conflict with the lower bound. This conjecture is controversial in its own right: for example, Bohmians tend to assume it is true \cite{pw4}, while Tony Leggett believes that it is not \cite{Leggett}, and it is also possible to adopt an intermediate position in which the branches coexist for a finite time. The latter possibility would be sufficient from our point of view, as long as this time is sufficiently long for the effects of $\epsilon$ to destroy one of the branches before they have a chance to recombine. Resolving this controversy would involve numerical simulations with too many particles for them to be currently feasible.

Another alternative would be to impose some external measurement, whether it would be the GRW model, or even Copenhagen collapse. As long as the measurement is external, we no longer need $\epsilon$ to be ``large enough" to account for it; since the ``finite lifetime" is strictly the consequence of ``finite epsilon", this means that we can now assert that the particle has an arbitrarily large lifetime but measurement ``kills it from the outside," so to speak. We would still, however, retain a finite $\epsilon$, both for philosophical purposes as well as to explain the direction of time. Finally, it can also be speculated that a finite $\epsilon$ arises out of the large timescale approximation to the GRW model \cite{GRWMensky}. But, again, for philosophical purposes we would like to believe in $\epsilon$ being finite even without the GRW mechanism, while the role of GRW is to magnify it, so to speak.

Let us now make some numerical predictions. First of all, Eq.\ \ref{BothDirectionsDecay} tells us that the lifetime of a particle ``to the future" from the given time is $2 \omega_k/\epsilon$. Since the lifetime ``to the past" is the same, the total lifetime is
\beq
\tau = \frac{4 \omega_k}{\epsilon}\;.
\eeq
This means that in the rest frame of any given particle its lifetime is 
\beq
\tau_{\rm rest} = \frac{4m}{\epsilon}\;,
\eeq
and then, since $\omega_k = \gamma m$, the Lorentz time dilation gives us
\beq
\tau = \gamma \tau_{\rm rest} = \frac{4 \gamma m}{\epsilon}
= \frac{4 \omega_k}{\epsilon}\;. \label{Lifetime}
\eeq
Let us now evaluate the relation between $\epsilon$ and the measurement precision. Suppose we are ``measuring" $\phi$ within a small four-volume $\delta v$. In this case, the ``weight" coming from the corridor is 
\beq
w(\delta\phi, \delta v) = \ee^{-\epsilon(\delta\phi)^2 \delta v/2}\;.\label{WidthOfVolume}
\eeq
This means that the standard deviation of $\phi$ inside that small volume $\delta v$ is 
\beq
\Delta\phi = \sqrt{\frac{2}{\epsilon\, \delta v}}\;. \label{DeltaPhi}
\eeq
On the other hand, Eq.\ \ref{Lifetime} implies that $\epsilon = 4\omega_k/\tau$, and if we substitute this into Eq.\ \ref{DeltaPhi} we obtain
\beq
\Delta\phi = \sqrt{\frac{\tau}{2 m \, \delta v}}\;.
\eeq
The lifetime $\tau$ is something that takes place in a ``large" scale, while the volume $\delta v$ is very small. Thus, $\tau$ is independent of $\delta v$, while $\Delta \phi$ is a function of the latter. We see therefore that $\Delta \phi$ is inversely proportional to $\sqrt{\delta v}$ (this can actually be seen by inspection of the ``weight" factor in Mensky's original formulation, even without reading anything else we have done), with a coefficient that is a function of the lifetime $\tau$ (this is where the non-trivial implication of our calculation comes into the picture).

\subsection*{5. Conclusion}

\noindent In this paper we have shown how two apparently unrelated and unsatisfactory aspects of QFT, the quantum measurement issue and the meaning of $\ii\epsilon$, are related. This result is quite interesting since the communities that would be interested in these two questions are very different: the former would be physicists and the latter mathematicians. So it seems to show that a rather ``technical" question that most physicists tend to push aside actually relates to something else which they find to be very interesting. What is especially interesting is that the only ``modification" needed to include ``quantum measurement" was shown to be \emph{part of} the QFT formalism as we know it, with a reinterpreted $\epsilon$, as long as the lower and upper bounds on $\epsilon$ mentioned earlier are compatible. Finally, we have made the non-trivial prediction of a ``finite" lifetime for scalar particles, and we have shown a dependence between their lifetime and the precision of field measurements; this part is a new prediction and doesn't follow from conventional QFT. 

To our knowledge this is the first time the ``measurement" follows from the path integral itself without the need for any other extra ingredients (Mensky's ``weight function" \cite{Proc,Mensky,Kent}, the GRW ``spontaneous localizations" \cite{GRW1985,GRW1986}, Bohmian beables \cite{pw1}, or any others), at least if the condition on the relationship between the upper and lower bounds on $\epsilon$ in Section 4 holds. In addition, under this equivalence the constant $\epsilon$ achieves a physical meaning which it previously lacked; the value of $\epsilon^{-1/2}$ is related to the classical scale, which can be estimated at least to some degree of accuracy.   

It is important to note that this paper treats only bosonic fields rather than 
fermionic ones. After all, fermions are defined using Grassmann numbers, which makes it unclear what the ontology of $\psi_{\rm cl}$ is, given that it isn't being integrated 
over. In principle, it is possible to address this issue by using a proposal on how to define Grassmann numbers outside the integral \cite{Grassmann}. However, it is not clear whether the restriction to the ``corridor" in Grassmann coordinates (as defined in that paper) would have physical implications similar to those found in the bosonic case, and the steps leading to the Gaussian function defining the corridor from the presence of $\epsilon$ in the Lagrangian do not go through as described here, since analytic functions of real Grassmann numbers are linear; whether this can be made to work in a modified derivation is left for future work. 

However, there is a much more conventional way to address the above concern about fermions. One can claim that we never ``see" an electron; we only ``see" the photons that an electron emits when it hits the screen. More generally, we can make the claim that we are only measuring bosonic fields and not fermionic ones (this is analogous to the assumption made in some pilot-wave models, such as Ref.\ \cite{pw4}, that only bosonic variables are beables). In terms of Feynman diagrams, this means that all of the incoming and outgoing lines are bosonic, while Fermions appear in loops. Now, when computing loops, we would still include $\ii \epsilon$. But this $\ii \epsilon$ will \emph{not} have the implications discussed in this paper; its purpose will be to avoid poles. But \emph{at the same time} the $\ii \epsilon$ that appears in the bosonic propagators will, in fact, do what we claimed it does. Consequently, bosonic fields will be measured while fermionic ones won't; yet we would get the information about fermions indirectly by measuring bosons.

\end{document}